\documentclass[11pt, prl, prstab,preprint,a4paper,showpacs,%
amsmath]{revtex4}
\usepackage{epsfig}
\usepackage{graphicx}
\usepackage{psfrag}

\newcommand{\sech}{\mathrm{sech}}

\newcommand{\sg}{\mathrm{sg}}

\newcommand{\parref}[1]{(\ref{#1})}

\newcounter{saveeqn}%

\newlength{\figwidth}
\begin{document}

\title{Quantum corrected electron holes}
\date{\today}

\author{A. Luque} \author{H. Schamel}\email{hans.schamel@uni--bayreuth.de}
\affiliation{Physikalisches Institut, Universit\"{a}t Bayreuth,
D--95440 Bayreuth, Germany}
\author{R. Fedele}
\affiliation{Dipartimento di Scienze Fisiche, Universit\`{a} di Napoli
and INFN Sezione di Napoli, Complesso Universitario di M. S. Angelo, Via
Cintia, I--80126 Napoli, Italy}

\begin{abstract}
  The theory of electron holes is extended into the quantum regime.
The Wigner--Poisson system  is solved perturbatively based in lowest order
on a weak, standing electron hole.  Quantum corrections are shown to lower
the potential amplitude and to increase the number of deeply trapped
electrons. They, hence, tend to bring this extreme non--equilibrium state
closer to thermodynamic equilibrium, an effect which can be attributed to
the tunneling of particles in this mixed state system.
\end{abstract}

\maketitle

Quantum plasmas\cite{Redmer} have recently attracted considerable
attention. Non--ideal, dense plasmas generated e.g. in the ultraintense
laser--solid interaction certainly belong to this category.  However, also
ideal, dilute plasmas --the addressee of this letter-- can exhibit a
quantum behavior.  One reason is that the miniaturization of today's micro-
and nano-electronic components has reached a level such that the system
length becomes comparable with the de Broglie wavelength, in which case
tunneling effects are no longer negligible \cite{Kluksdahletal}. Also,
states in combined traps attained by particles and their anti--particles
used to form anti--hydrogen may be modelled by a plasma having quantum
features \cite{Holzscheiter}.   Other examples can easily be found, and
some of them will be mentioned further below. Hence, classical transport
models will unlikely be sufficient to describe the plasma behavior in such
devices adequately. Generally the main focus in this respect is the
collective charge
carrier transport which has been shown \cite{KornSchamel12} to be increased
by the presence of coherent structures such as hollow phase space vortices
\cite{Schamel79a86}.  The latter are also called holes in plasma physics
\cite{KornSchamel12, Schamel79a86, Saeki, SchamelBujarbarua} because
of their associated
density depression exhibiting a remarkable robustness and stability
\cite{Schamel82ab}.

Before studying the quantum corrections to such holes
let us review some further dilute many particle systems and related
disciplines that reveal quantum aspects.
 Charged--particle beams in particle
accelerators are typically dilute systems, so quantum effects are usually
disregarded. However, a spectrum of phenomena, which recently became more
and more important, reveals the existence of several quantum aspects of
beam physics connecting the physics of particle accelerator with the
frontiers of several disciplines, such as (for instance) plasma physics,
radiation beam physics, astrophysics, mesoscopic and condensate physics
\cite{Chen1,Chen2}. Most of these phenomena introduce a sort of quantum
correction to the leading classical behavior of the system. For example,
quantum excitation \cite{Sands} plays a role for the long--term stability
of longitudinal electron beam dynamics in the high--energy circular
accelerating machines while the Sokolev--Ternov effect \cite{SokolevTernov}
of spin polarization of electron and proton beams is a manifestation, at
the macroscopic level, of the single quantum nature of the beam particles.
Numerical phase space investigations based on tracking with the quantum map
have shown that quantum corrections can substantially affect the particle
beam trajectories in the vicinity of the separatrix \cite{Heifets}.

Recently quantumlike methodologies \cite{Fedele2DeMartino} have been
applied to a number of classical physical situations, in which $\hbar$
is replaced by another characteristic parameter of the
problem considered. For instance, they have been
applied to accelerator physics \cite{Fedele3a3bAnderson}, to
plasma physics
\cite{Shukla}, to surface gravity wave physics \cite{Alberetal} and to
nonlinear optics
\cite{Leontovichetal} in an attempt to
describe linear and nonlinear problems of 
the dynamics of beams and
large amplitude wavepackets.

In principle, all these problems can be formulated, in the configuration
space, in terms of a system of Zacharov equations, i.e. nonlinear
Schr\"{o}dinger--like equation coupled with one (or more) equation(s) taking
into account the reaction of the environment. The corresponding phase space
description is the one provided by the Wigner--Moyal quasidistribution
\cite{Wignerall}
 whose evolution equation, the von Neumann
equation, plays the role of a kinetic--like equation associated with the
system.

Analytically, the framework under which Wigner--Moyal quasidistributions
have been mostly considered so far is that of particles interacting with a
given external e.g. parabolic potential to analyze coherent and squeezed
states. Furthermore, a quantum-like phase space analysis of a
paraxial--charged--particle
beam transport, travelling through a quadrupole-like device with small
sextupole and octupole aberrations, has been carried out showing a
satisfactory agreement with the results of the standard tracking
simulations \cite{Fedele4}
 and, consequently, the suitability of using the quantum phase--space
formalism in particle accelerators. This has been done within the framework
of the thermal wave model \cite{Fedele3a3bAnderson}.
Quantum-like corrections involved
in the von Neumann equation have been discussed for paraxial beams of both
particles and radiation and compared with the standard classical
description \cite{Fedele55a}.

Particles in quantum plasmas moving in their own, self--consistent
potential, on the other hand, have not been given much attention so far. An
exception are self--consistent but linearized solutions of the
Wigner--Poisson system, dealing with quantum corrections to Landau damping
of Langmuir waves \cite{Klimontovich60} or to the two--stream instability
by means of the Nyquist method \cite{Haas01}, and the self--consistent
linearized solution of the Wigner--Moyal kinetic--like equation for
Langmuir wavepackets coupled with the ion--acoustic wave equation
\cite{Fedele02}. In particular, the Wigner--Moyal kinetic--like description
is suitable for describing the Benjamin--Feir instability (modulational
instability) as well as predicting the stabilizing effect of a sort of
Landau damping. It is well known that the latter cannot be shown in
configuration space, where the the system is usually described by the
Zakharov equations. By using the pure state formalism, a Landau--type
damping has been shown for the longitudinal dynamics of both
charged--particle coasting beams and e.m. wavetrains in high--energy
circular accelerators and nonlinear media, respectively \cite{Fedele66a}. A
similar approach has been extended (mixed state formalism) to ensembles of
partially--incoherent waves in different physical situations
\cite{Fedele02, Hall, Onorato}.

 The aim of this paper is to describe an electron--ion unmagnetized
plasma, in which, on the basis of the experimental evidences as well as on
theoretical and numerical investigations mentioned above, the quantum
nature of the particles is not disregarded. However, it is taken into
account only as a weak (perturbative) effect in comparison to the leading
classical behavior of the system. Together with the weak quantum effect,
we take into account the usual classical electrostatic collective plasma
effects coming from the standard meanfield approximation of the coulombian
interaction, in such a way that our system is described by a set of coupled
equation comprising the von Neumann equation for the Wigner--Moyal
quasidistribution and Poisson's equation. Hereafter, we will refer to
this system of equations as WP--system (Wigner--Poisson system). Our goal
is to find a self consistent solution of the WP--system to the lowest order
of the quantum correction. Before formulating our problem in detail, it is
worth mentioning some further considerations starting from the classical
case.

Classically, electron and ion holes are nonlinear, stationary solutions of
the Vlasov--Poisson system (VP--system) being omnipresent structures in
many driven, collisionless plasmas.  They are found in one or other variant
in the laboratory \cite{Saeki, Moody, Petraconi}, in particle accelerators
\cite{Colestock, Koscielniak}, in the laser--plasma interaction
\cite{Montgomery2001} and in the extraterrestrial space
\cite{Omura96Mozer97}. Analytical
solutions, applicable to kinetic hole structures found in particle
accelerators, have been presented in \cite{SchamelAcc}.

Generally speaking, the fundamental role of holes arises from the fact that
they can nonlinearly destabilize a plasma even in linearly stable
situations, namely if they posses a negative energy character
\cite{GriessmeierSchamelPlasma, GriessMeierLuqueSchamel}.  There is hence
accrued interest to extend the studies of holes into the quantum
domain, 
which means that the WP--system has to be employed instead.

In recent numerical studies \cite{Haas00} a multistream model for a
current--driven quantum plasma has been applied. Signatures of coherent
hole structures appear in the simulation of a statistical mixture of many
pure states, with each wave function obeying the Schr\"{o}dinger--Poisson
system. That this system is equivalent to the mixed state WP--system
has been shown in
\cite{Markowich}. To the best of our knowledge, an analytic
self--consistent nonlinear solution of the WP--system is still missing in
the literature.

In this letter we present a first rigorous nonlinear self--consistent
solution of the WP--system assuming weak nonlinearity and proximity to the
classical VP--system.  First we shall refer to the classical e--hole and
then study quantum corrections, for which a self--consistent
solution is derived.

\label{eholes}
We are first investigating a standing, classical e--hole which is the
simplest inhomogeneous stationary solution of the VP--system for a plasma
with immobile ions:
\begin{subequations}
\begin{equation}
        \left[v\partial_x + \phi'(x)\partial_v\right]f(x,v) = 0,
        \label{vlasov}
\end{equation}
\begin{equation}
        \phi''=\int dv f(x,v) - 1 := n_e(x) - 1,
        \label{poisson}
\end{equation}
\end{subequations}
where $f(x,v)$ is the distribution function of electrons in phase
space.  Here space, velocity and density
are normalized by the electron Debye length
$\lambda_D = (kT_e/4\pi n_0 e^2)^{1/2}$, the electron thermal
velocity $v_{th} = (kT_e/m_e)^{1/2}$ and the homogeneous electron density,
$n_0$, where $m_e$ is the mass of the electrons and $T_e$ is
their temperature.

In thermal equilibrium, the plasma adopts a homogeneous state
with a Maxwellian distribution in velocities
$f_M(x,v) = \frac{1}{\sqrt{2\pi}}\exp\left(-v^2/2\right)$.
Referring to the potential method \cite{Schamel7275}, we solve
\parref{vlasov} by
\begin{equation}
        f(x,v) = \frac{1}{\sqrt{2 \pi}}\left[
                \exp(-E)\theta(E) + \exp(-\beta E)\theta(-E)\right],
        \label{fclas}
\end{equation}
where $E=v^2/2 - \phi$ represents the single electron energy.  The
separatrix $E=0$ in phase space distinguishes the free ($E>0$) from
the trapped ($E<0$) electron population. Note that $f$ in
\parref{fclas}, reduces to $f_M$ as $\phi \to 0$.  The electron
density in the weak amplitude regime can be written as a half power
expansion of $\phi$~\cite{Schamel79a86, Schamel7275},
\begin{equation}
        n_e(\phi) = 1 + \phi -
        \frac{4(1-\beta)}{3\sqrt{\pi}}\phi^{3/2} + O(\phi^2).
        \label{n_e}
\end{equation}
Defining the pseudo--potential as $-V'(\phi) := n_e(\phi) - 1$,  we find
from \parref{n_e} with $V(0)=0$
\begin{equation}
        -V(\phi) = \frac{\phi^2}{2} \left(1 -
                 \frac{16(1 - \beta)}{15\sqrt{\pi}}\sqrt{\phi}\right).
        \label{V(phi)}
\end{equation}
It has to fulfill two necessary conditions: (a) $V(\phi) < 0$ in $0 <
\phi < \psi$ and (b) $V(\psi) = 0$, where $\psi$ is the amplitude of
the perturbation in the potential, which is assumed to be small, $\psi \ll
1$.  From (b) we arrive at $-\beta = 15\sqrt{\pi}/16\sqrt{\psi} - 1 \approx
15\sqrt{\pi}/16\sqrt{\psi} \ll 1$. Therefore, $\beta$ has to be a large
negative number, corresponding to a depletion of the distribution in the
trapped particle range.  On the other hand, equation \parref{V(phi)} allows
us to integrate Poisson's equation \parref{poisson}, and we obtain the
bell--shaped electrostatic potential:
\begin{equation}
        \phi(x) = \psi \sech^4\left(\frac{x}{4}\right).
        \label{phiclas}
\end{equation}
Note that other electrostatic structures such as propagating electron
holes, ion holes or periodic nonlinear waves (cnoidal waves)
can also be found by appropriate extensions of this method
\cite{Schamel79a86, SchamelBujarbarua, Schamel7275}.

  To study quantum corrections, we start with Wigner's
quasidistribution which satisfies the time independent von Neumann or
quantum Liouville equation\cite{Wignerall}:
\begin{equation}
        v\partial_x f + \frac{1}{i\varepsilon}
        \left[\phi\left(x + \frac{i\varepsilon}{2}\partial_v\right) -
              \phi\left(x - \frac{i\varepsilon}{2}\partial_v\right)\right]f = 0,
        \label{vonNeumann}
\end{equation}
where $\varepsilon$ is the dimensionless Planck's constant
$\varepsilon = \frac{\hbar}{m_e v_{th} \lambda_D}$.

  Assuming that $\varepsilon$ is small, i.e. the quantum
effects appear only as corrections to the classical solution, we can
perform a power expansion of the potential operator $\phi\left(x {\pm}
\frac{i\varepsilon}{2}\partial_v\right)$ which we insert into
\parref{vonNeumann}.  All even terms cancel out and we get up to the
third order
\begin{equation}
        v \partial_x f + \phi'(x)\partial_v f -
                \frac{\varepsilon^2}{3!4}\phi'''(x)\partial_v^3 f = 0,
        \label{vonNeumann2}
\end{equation}
which is the equation we have to couple with \parref{poisson}.

  As we keep terms up to $O(\varepsilon^2)$, we will look for corrections of
the same order in the potential and in the distribution function, $f =
f_0 + \varepsilon^2 f_1$, $\phi = \phi_0 + \varepsilon^2 \phi_1$, with $f_0$
and $\phi_0$ representing now \parref{fclas} and \parref{phiclas}
respectively. Inserting this ansatz into \parref{vonNeumann2} and
\parref{poisson} neglecting again terms of $O(\varepsilon^4)$
we find
\begin{subequations}
\begin{equation}
        \left[v \partial_x + \phi_0'(x) \partial_v\right]f_1
        = -\phi_1'(x)\partial_v f_0 + \frac{1}{3!4}\phi_0(x)'''
        \partial_v^3 f_0,
        \label{vonNeumann3}
\end{equation}
\begin{equation}
        \partial^2_x \phi_1(x) = \int dv f_1.
        \label{poisson2}
\end{equation}
\end{subequations}

  By defining $g(x, v) := f_1 + \phi_1 \partial_E f_0$, we can reduce
equations \parref{vonNeumann3} and \parref{poisson2} to the somewhat
simpler system
\begin{subequations}
\label{14}
\begin{equation}
        \left[v\partial_x + \phi_0'(x)\partial_v\right]g =
                \frac{1}{3!4}\phi_0'''(x)\partial_v^3f_0 =: h(v,x),
        \label{14a}
\end{equation}
\begin{equation}
        \phi_1''(x) + V''(\phi_0)\phi_1(x) =
                \int_{-\infty}^{+\infty}dv\,g(x,v),
        \label{14b}
\end{equation}
\end{subequations}

Now it is convenient to switch into a new set of variables defined by
$\xi = x$, $E = \frac{v^2}{2} - \phi_0(x)$, $\sigma=\sg(v)$
and rewrite $h(x,v) = H(\xi,E,\sigma)$, $g(x,v) = G(\xi, E, \sigma)$.
With these variables, \parref{14a} becomes
$\partial_\xi G(\xi, E, \sigma) = H(\xi, E, \sigma)/v(\xi,E,\sigma)$,
whose general solution is
\begin{equation}
        G(\xi,E,\sigma) = G(0,E,\sigma) +
          \int_0^{\xi}d\xi'\,\frac{H(\xi',E,\sigma)}{v(\xi',E,\sigma)},
        \label{fullG}
\end{equation}
where $v(\xi,E,\sigma) = \sigma \sqrt{2\left[E+\phi_0(\xi)\right]}$.
Therefore, in order to find $G$ we only have to
integrate $H(\xi,E,\sigma)/v(\xi, E, \sigma)$ along the classical
particle trajectories given by $E=const.$.  In this expression we have
chosen the lower integration limit as $\xi=0$ because this is the only
point which is reached by all trajectories (see below).  Note that a
trapped particle will move along a closed, bounded trajectory around
the origin in phase space.

  Now we need to replace $H(\xi,E,\sigma)$ by its full expression.
Denoting $f$ in Eq.~\parref{fclas} as $f_0(E)$ we get by differentiation
\begin{eqnarray}
        \nonumber
        \partial_v^3f_0 & = & \frac{1}{\sqrt{2\pi}}
          v\left\{\left[3-2 (E+\phi_0)\right]e^{-E}\theta(E)
          + \beta^2\left[3-2\beta (E+\phi_0)\right]e^{-\beta E}\theta(-E)
                \right.\\
                \nonumber & & \left.
          - \left[3(1-\beta) -2\phi_0(1-\beta^2)\right]\delta(E)
          - 2(E+\phi_0)(1-\beta)\delta'(E)\right\} \\
         & =: & \frac{1}{\sqrt{2\pi}} v(\xi,E,\sigma) \Omega(\xi,E).
\end{eqnarray}

For positive energies, we can follow the trajectories up to any
$\xi$ in \parref{fullG} and, assuming that the correction vanishes
at $\xi\to{\pm}\infty$, we arrive at
\begin{equation}
        G(0,E,\sigma) =
           \int_{-\infty}^{0}d\xi'\,\frac{H(\xi',E,\sigma)}{v(\xi',E,\sigma)}
          =\frac{1}{3!4\sqrt{2\pi}}\int_{-\infty}^{0}d\xi'\,
                \phi_0'''(\xi')\Omega(\xi',E).
\end{equation}
Note that this expression does no longer depend on $\sigma$. For negative
energies $G(0,E,\sigma)$ is not determined by such a procedure but,  due to
the symmetry of the problem we can assume that it will also be
$\sigma$--independent.  On the other side, we can always extend the
integration  of \parref{fullG} to $-\infty$ for negative energies also as
long as we change the integration constant. Therefore we have, for any $E$,
\begin{equation}
        G(\xi,E,\sigma) = G(\xi,E) := \frac{1}{3!4\sqrt{2\pi}}\left[G_0(E) +
        \int_{-\infty}^{\xi}d\xi'\,\phi_0'''(\xi')\Omega(\xi',E)\right],
        \label{G}
\end{equation}
with $G_0(E) = 0$ for $E > 0$.

  Note that $\partial_E f_0$ is discontinuous at $E=0$.  Therefore,
$G(\xi, E)$ does not have a definite value at the separatrix.  Our
approach will be to solve \parref{14} for positive and negative
energies separately and then put both solutions together imposing the
continuity of $f_1$ at the separatrix.

  In order to integrate \parref{G}, we consider these two different cases:
\begin{enumerate}
\item  For $E > 0$ we have $G_0(E)=0$ and $\Omega(\xi,E) = \left[3 -
2(E+\phi_0(\xi))\right]e^{-E}$.
The integral \parref{G} can be performed analytically to yield
\begin{eqnarray}
        \nonumber
        G(\xi,E) & = & \frac{1}{3!4\sqrt{2\pi}}
                \left[\phi_0'(\xi)^2 + \left(3 - 2E -
                2\phi_0(\xi)\right)\phi_0''(\xi)\right]e^{-E}.
        \label{G(E>0)}
\end{eqnarray}
\item If $E<0$ we must take $G_0(E)$ into account.  In this case
$\Omega(\xi,E) = \beta^2\left[3 -
2(E+\phi_0(\xi))\beta\right]e^{-\beta E}$
and \parref{G} reads
\begin{equation}
        G(\xi, E) = \frac{1}{3!4\sqrt{2\pi}}\left\{
               G_0(E) +
                \left[\beta\phi_0'(\xi)^2 + \left(3 - 2\beta E -
                2\beta\phi_0(\xi)\right)\phi_0''(\xi)\right]
                        \beta^2 e^{-\beta E}\right\}.
        \label{G(E<0)}
\end{equation}
\end{enumerate}

  The continuity of $f_1$ is now imposed to determine $G_0(E)$.
As $f_1 = g - \phi_1\partial_E f_0$, the discontinuity  of $g(x,v) =
G(\xi, E)$, namely $\Delta G := G(\xi,0^+) - G(\xi,0^-)$ should be
equal to $\phi_1\Delta(\partial_E f_0)$.  Since it holds
$\Delta(\partial_E f_0) = (\beta - 1)/\sqrt{2\pi}$, we get
\begin{eqnarray}
        \nonumber
        \Delta G = &\frac{1}{3!4\sqrt{2\pi}}&\left\{
                \phi_0'(\xi)^2+\left(3-2\phi_0(\xi)\right)\phi_0''(\xi)
                \right.\\
                & &
                \left.
                -\beta^2\left[\beta\phi_0'(\xi)^2+
                              \left(3-2\beta\phi_0(\xi)\right)
                                \phi_0''(\xi)\right] - G_0(0)\right\}.
        \label{DeltaG}
\end{eqnarray}
Then we can find $\phi_1(\xi)$ as
$\phi_1(\xi) = \Delta G / \Delta(\partial_E f_0) = \sqrt{2\pi}\Delta
G/(\beta-1)$ with $\Delta G$ given by \parref{DeltaG}.
Moreover, as we impose $\phi_1({\pm}\infty) = 0$, we know that
$G_0(0)=0$.  Hence, we obtain $\phi_1(\xi)$.
\begin{figure}
\includegraphics[width=\figwidth]{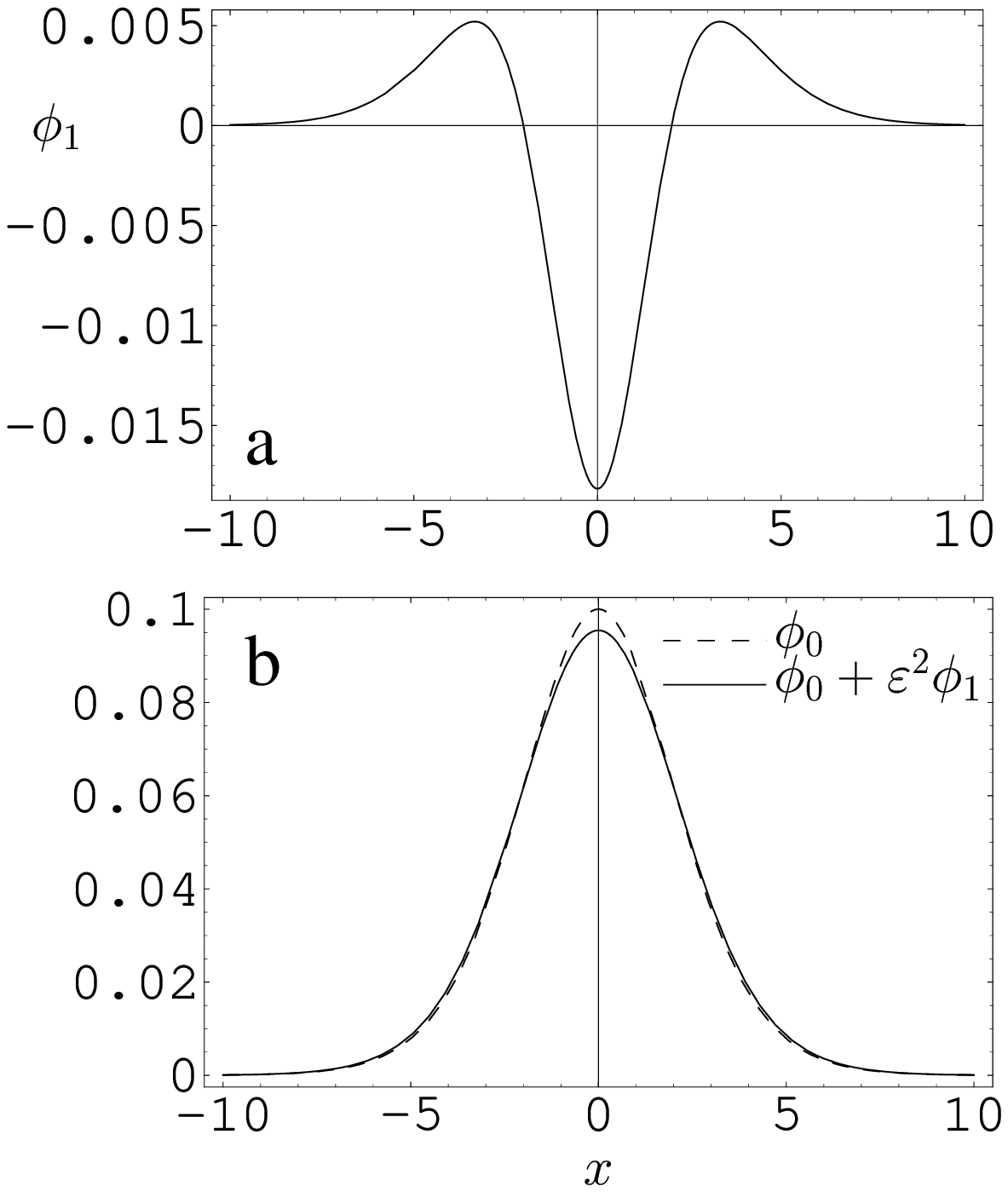}
\caption{\label{Fig1}
a) Correction to the potential for $\psi = 0.1$; b) the classical and
the quantum corrected potential.}
\end{figure}
Figure \ref{Fig1}a shows $\phi_1(\xi)$ for $\psi=0.1$ and
Fig.~\ref{Fig1}b  represents the corrected  potential.  For reference,
also the unperturbed potential is drawn.  We see that the potential
experiences a reduction as a result of quantum correction.

To determine $G_0(E)$ for all negative energies we go back to \parref{14b}.
As we have already determined $\phi_1(x)$, the left hand side is now given.
It is convenient to write it in terms of $\phi_0$.  To do this we note that
all derivatives of $\phi_0$ can by expressed by $\phi_0$ itself via
$\phi_0'(\xi) = -\phi_0(\xi)
          \left(1 - \sqrt{\phi_0(\xi)/\psi}\right)^{1/2}$,
$\phi_0''(\xi) = \phi_0(\xi)
          \left(1 - 5\sqrt{\phi_0(\xi)/\psi}/4\right)$ and also
$V''(\phi_0)=-\left(1 -
        15\sqrt{\phi_0/\psi}/8\right)$.
Inserting these expressions into \parref{14b} we find an
expression for its left hand side as a function of $\phi_0$ which we
call $L(\phi_0)$. An
explicit form of this function is too long to be included in this
letter and will be presented elsewhere\cite{LuqueSchamelFedele}.

On the other hand, the right hand side of \parref{14b} can be written
as
\begin{equation}
        \int_{-\infty}^{+\infty}dv\,g(x,v) =
          \sum_\sigma \sigma\int_{-\phi_0(\xi)}^{\infty}dE\,
                            \frac{G(\xi,E)}{v(\xi,E,\sigma)}
        = 2\int_{-\phi_0(\xi)}^{\infty}dE\,
                            \frac{G(\xi,E)}{v(\xi,E,1)}.
        \label{PrevIntG}
\end{equation}
Making use of \parref{G} we can reduce \parref{PrevIntG} to
\begin{equation}
\begin{split}
        \int_{-\infty}^{+\infty}dv\,g(x,v) = R(\phi_0) +
                \frac{1}{3!2\sqrt{2\pi}}
                \int_{-\phi_0(\xi)}^{0}dE\,\frac{G_0(E)}{v(\xi,E,1)}
        \label{IntG}
\end{split}
\end{equation}
Where $R(\phi_0)$ represents a known function that can be
obtained analytically\cite{LuqueSchamelFedele}.

\begin{figure}
\includegraphics[width=\figwidth]{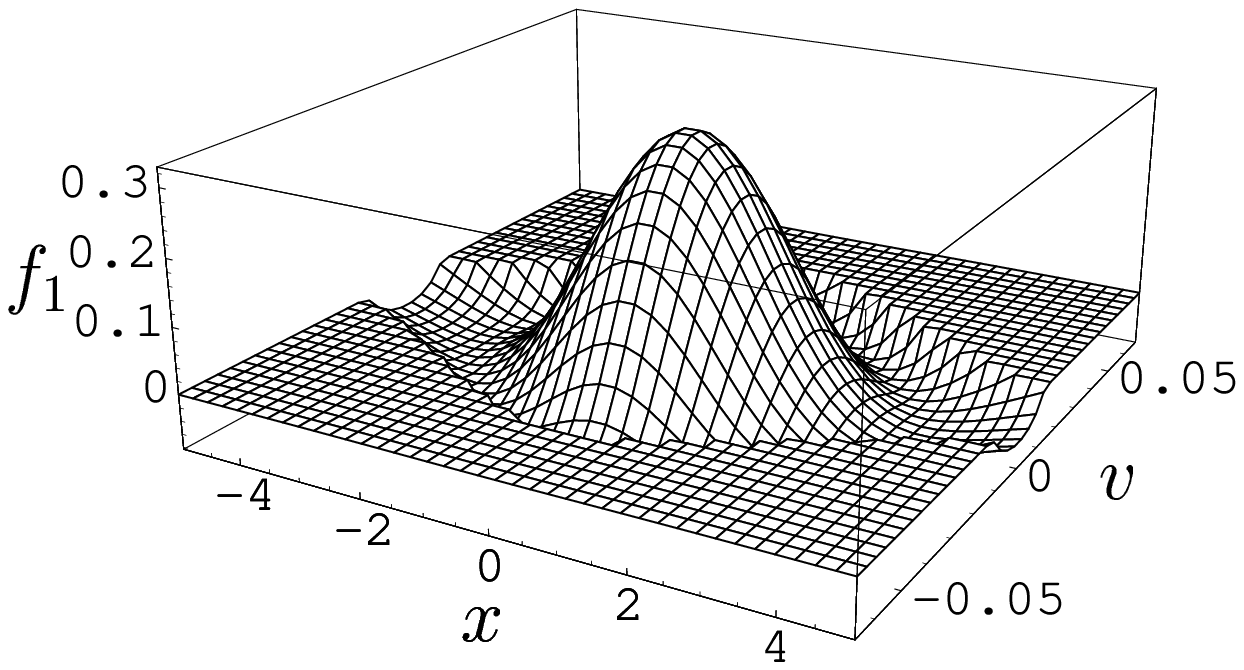}
\caption{\label{Fig3}
Correction of the distribution function $f_1(x,v)$.}
\end{figure}

\begin{figure}
\includegraphics[width=\figwidth]{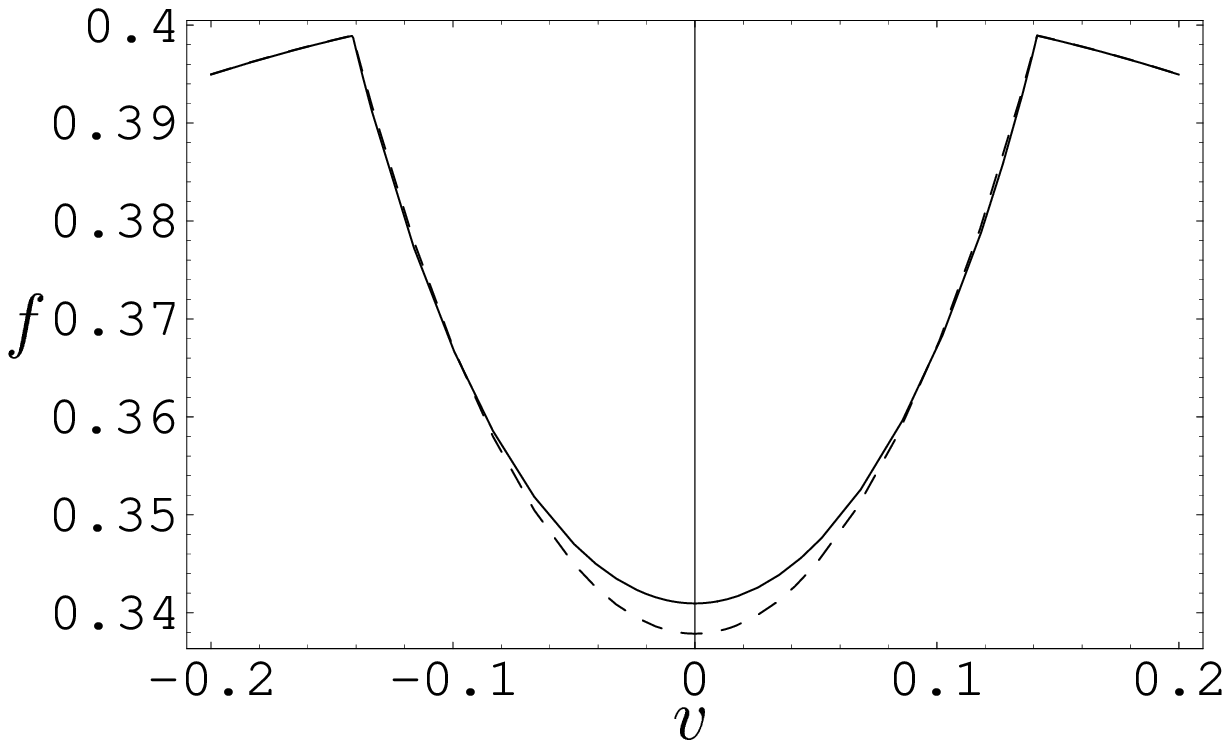}
\caption{\label{Fig4}
Corrected distribution function $f=f_0+\varepsilon^2 f_1$ at $x=0$, for
$\psi=0.01$, $\varepsilon=0.1$.  The dashed line represents the
original (unperturbed) distribution function.}
\end{figure}

To perform the remaining integral of \parref{IntG}, we make a half power
expansion of $G_0(E)$:
\begin{equation}
        \frac{1}{3!2\sqrt{2\pi}}G_0(E) =
        a_{1/2}|E|^{1/2} + a_{1}|E| + a_{3/2}|E|^{3/2} +\dots
\end{equation}
With this ansatz, we have
\begin{equation}
        \frac{1}{3!2\sqrt{2\pi}}
        \int_{-\phi_0(\xi)}^{0}dE\,\frac{G_0(E)}{v(\xi,E,1)} =
                \sqrt{\phi_0}\sqrt{\frac{\pi}{2}}\sum_{n=1}^{\infty}
                \frac{\Gamma\left(1 + \frac{n}{2}\right)}
                     {\Gamma\left(\frac{3}{2} + \frac{n}{2}\right)}
                     a_{n/2}{\phi_0^{n/2}}.
\end{equation}
And we can finally reduce \parref{14b} to
\begin{equation}
        L(\phi_0) = R(\phi_0) + \sqrt{\phi_0}\sqrt{\frac{\pi}{2}}
                \sum_{n=1}^{\infty}
                \frac{\Gamma\left(1 + \frac{n}{2}\right)}
                     {\Gamma\left(\frac{3}{2} + \frac{n}{2}\right)}
                     a_{n/2}{\phi_0^{n/2}}.
\end{equation}
Therefore, if we define $\rho(t) := \frac{1}{t} \left(L(t^2)-R(t^2)\right)$,
we can find all $a_{n/2}$ as
\begin{equation}
        a_{n/2} = \frac{1}{n!}\sqrt{\frac{2}{\pi}}
                \frac{\Gamma\left(\frac{3}{2} + \frac{n}{2}\right)}
                     {\Gamma\left(1 + \frac{n}{2}\right)}
                \left.\frac{d^n\rho(t)}{dt^n}\right|_{t=0}
\end{equation}
With this expression for $a_{n/2}$ we can sum $G_0(E)$ and then find
$G(\xi,E)$ and $f_1$.  The correction of the distribution function,
$f_1(x,v)$ is plotted in Fig.~\ref{Fig3},
while the final, corrected distribution function $f=f_0 +
\varepsilon^2 f_1$ is represented at fixed $x$ in
Fig.~\ref{Fig4}.

  We clearly recognize a partial filling of the phase space within the
separatrix being maximum at the hole center.  An interpretation may be
given in terms of refraction or tunneling: in the classical solution
nearby its separatrix, the region of untrapped electrons is populated
stronger than that of trapped electrons.  In the quantum domain when
tunneling becomes effective this gives rise to a net influx of
particles resulting in a less dilute distribution of trapped
electrons. This interpretation of the collective particle behavior,found
analytically,matches well with the numerical findings of 
Ref.~\onlinecite{Heifets},
according to which the quantum corrections affect the particle
trajectories in the vicinity of the separatrix (see also
Ref.~\onlinecite{Fedele4}).

  We, therefore, conclude that the overall effect of a quantum
correction to a classical e--hole is the tendency of the system to reduce
the coherent excitation by both a diminution of the amplitude and a partial
filling of the trapped particle region by refraction (tunneling), bringing
the system closer to the thermal state.

  Open questions are how these semiclassical corrections are modified
in case of finite amplitudes $\Psi \gtrsim O(1)$, of finite quantum
corrections $\varepsilon \gtrsim O(1)$, of hole propagation $v_0 > 0$,
of nonlocality of structures such as periodic wave trains, some of
which will be addressed in our forthcoming
publication\cite{LuqueSchamelFedele}.


\label{aknowledgements}
  This work was performed under the auspices of the DAAD (Deutscher
Akademischer Austauschdienst) and CRUI (Conferenza dei Rettori delle
Universita Italiane) in the framework of the research program ``VIGONI.''

%\bibliography{../plasma,wigner}

\end{document}